# On the mathematical representation of nonlinearity

*Emanuel Gluskin*

Electrical Engineering Department,
Academic Technological Institute, 52 Golomb St., Holon 58102, Israel, and
Electrical Engineering Department, Ben-Gurion University, Beer-Sheva 84105.

**Abstract**: The suggestion of writing, for some problems, nonlinear state equations not as $d\mathbf{x}/dt = \mathbf{F}(\mathbf{x},\mathbf{u},t)$, but as $d\mathbf{x}/dt = [A(t,\mathbf{x})]\mathbf{x} + [B(t,\mathbf{x})]\mathbf{u}(t)$, which is more "constructive", is considered supported by arguments related to: the axiomatization of system theory, the classification of switched circuits as linear and nonlinear, the use of nonlinear sampling for measuring frequency without clock, the consideration of an ensemble of colliding particles in which the establishing of the chaotic movements is seen as a result of the same kind of nonlinearity as in electronic switching circuits, the modeling of a liquid medium as a "system" whose structure is directly influenced by the "input", and some other problems.

## 1. Introduction

It is suggested in the remarkable pedagogical work [1] to start the laboratory study of electrical circuits by electrical engineering students from nonlinear, and not linear, circuits, and the rationale for teaching nonlinear circuits before ones linear is articulated in the capital source [2] too. Obviously, an acquaintance with nonlinear circuits before linear ones does not encourage one to just grammatically define/see "nonlinear" as "not a linear one" (which some students could erroneously understand as "not a good one"), but as something *constructively defined and described*. This position of [1,2] seems to us to be important not only pedagogically, but also axiomatically and we deal with the constructive definition of nonlinearity, suggesting some associated notations relevant to system classification, and develop an outlook, via system concepts, on some physical phenomena. The physical aspects contribute to better understanding of the system concepts and to making basic system theory become a part of one's general education.

Singular systems are one of our main focuses, and among such systems, switched systems are mentioned most often. Definitions of linear and nonlinear switched systems are unusually close because the very fact of switching has to be considered first of all, and the nonlinearity is then directly seen from the nature of the time-functions that control the switching. In the nonlinear case, this "nature" is a map $\mathbf{x}(t)$ → **t\*** of the state-variables $\{x_p(t)\}$ of a system on the switching instances $\{t_k\}$ that are important system parameters. This is inevitably done in the *structural terms* of the elements being switched. This approach is started in [3-5], but contrary to the original line of [3] which is more oriented towards a designer of switched circuits, here we are concerned with logical basics, and cover points that should be more interesting for a theoretician.

Some seemingly simple concepts had to be revisited; in particular, that of input (port) function. Without noting the freedom that modern switched systems give to the input connections, and thus to the very concept of "input", one can have difficulty in seeing possible nonlinearity even of some very simple systems/equations.

The main equational frame of the work is as follows.

Usually, nonlinear state-equations are written as (*t* is time and **x** is the vector of state-variables)

$$d\mathbf{x}/dt = \mathbf{F}(\mathbf{x},\mathbf{u},t) \qquad (1)$$

(dim **F** = dim **x**), but we use equations in the form that is closer to that of linear state-equations (for instance many classical Switched Capacitor Circuits/Filters are described as (2)),

$$d\mathbf{x}/dt = [A(t)]\mathbf{x} + [B(t)]\mathbf{u}(t) , \qquad (2)$$

i.e. as

$$d\mathbf{x}/dt = [A(t,\mathbf{x})]\mathbf{x} + [B(t,\mathbf{x})]\mathbf{u}(t) . \qquad (1a)$$

The nonlinearity in (1a) is thus seen via the influence of **x** (and we can call such a system an "**x**-system") on the "structure" understood as in linear systems, which makes a difference with respect to (1) where analytical dependencies are on the first place.

Of course, there is no suggestion of being detached from (1) in any case. For instance the example of Section 4.3 with a hardlimiter nonlinearity would require (see also [3]) [A(**x**)] to have a pole, which is unusual. However in the very important applications for switched systems, when a *nonlinear* system "jumps" at some-time instants *defined by* **x**(*t*), from one LTI system to another, form (1a) certainly has an advantage in understanding the system.

There also can be nonsingular systems, as e.g. in the physical analogy in Section 5.1, when (1a) is more adequate than (1). In particular, a linearization is obtained in this analogy very naturally, as the transfer from (1a) to (2) by substituting in (only) the matrices a certain approximating vector function $\mathbf{x}_o(t)$ instead of **x**(*t*).

Starting from Section 3, we shall often speak only about [A]. If **u** ≠ 0, then similar things are meant to apply to [B]. However, for nonlinearity of (1a) it is also sufficient that only [B] depends on **x**; some such systems of the second order are considered in [6,7] (d*x*/d*t* = P*x* + B(*x*,*t*), in the notations of [6]), but only as regards limit cycles.

The nonlinear non-autonomous systems have a very good interpretation here; *a system is nonlinear if it's structure is influenced by the input*. This influence can be indirect, expressed as [A(**x**(**u**))], or direct, expressed as [A(**u**)]. In the latter case, we speak about "**u**-systems" that are treated in Section 2. The latter section completes also some other knowledge about nonlinear equations that we actually need here.

The main line starts in Section 3 and 4 from **x**-systems that are most common nonlinear realization.

Section 5 shows that our system approach can be useful for some physical problems.

Sections 6 is devoted to consideration of another important singular operation, -- *sampling*, and for the nonlinear case an analogy to the procedure defining Lebesgue's integral is considered. Section 7 extends the topic of nonlinear sampling to an application to spectrum analysis.

Section 8 overviews the main results and conclusions adding some final comments.

Appendix 1 is a quite pragmatic completion to the discussion of switched systems in the main text, and Appendix 2 is devoted to some purely logical points that should interest a theoretician.

For some problems below it is useful to somewhat generalize the Small Theorem of [4] in which only the *scaling test* of linearity **u**→*k***u** (*k* is constant) is mentioned. This theorem says that if with increase in *k* starting from zero, a switching appears or disappears (or, in a periodic process, the density of the switching instants is changed), then the system is nonlinear. The proof uses the facts that any switching necessarily causes a singularity, and that two waveforms (shapes) having different numbers (or densities) of the points of singularity cannot be proportional each other, i.e. **u** → *k***u** cannot be followed by **x**→ *k***x**, as linearity requires. It is possible, as the point of the generalization, to also involve the *linear test of additivity*, by noting that if **u₁** + **u₂** causes a new switching that was not caused by either **u₁** or **u₂**, then the system is nonlinear. Indeed, if neither of the associated **x₁** and **x₂** includes a singularity at an instant, then their sum also cannot include it. This means that **u₁₍₂₎** → **u₁** + **u₂** cannot be followed, under the theorem's condition, by **x₁₍₂₎** → **x₁** + **x₂**, i.e. the system is nonlinear. It is also important for relevance of the theorem that switching is a clearly registerable phenomenon/operation. The point of measurement is important both axiomatically and practically (Appendix 2), and according to our actual perception of nonlinearity as something constructively defined, just "not a linear one" is not yet "nonlinear".

## 2. The "u-systems" and the concept of "given input"

It is the flexibility and the structural freedom existing in the design of modern electronics switching systems (e.g. Fig. 2 in [4] includes such a possibility), which cause us to consider "**u**-*systems*", in which the input can directly influence the structure:

$$d\mathbf{x}/dt = [A(\mathbf{u})]\mathbf{x} + [B(\mathbf{u})]\mathbf{u}(t) , \qquad (3)$$

or

$$d\mathbf{x}/dt = [A(\mathbf{u})]\mathbf{x} . \qquad (4)$$

For such an equation, the map

$$\mathbf{u} \to \mathbf{x}$$

that is realized by the associated system is *nonlinear*. Indeed, the scaling test of linearity (let us for simplicity always take zero initial conditions), i.e. the map

$$k\mathbf{u} \to k\mathbf{x} \quad (k \text{ is constant}) \qquad (5)$$

is *not* allowed by (3) or (4), obviously. One can find that this seems to contradict with the fact that since **u**(*t*) is given, [A(**u**)] and [B(**u**)] are also given, as [A(*t*)] and [B(*t*)] in (2), and though it is just a particular case of (4), it is not easy for many to accept that the equation

$$\frac{dx}{dt} + a(t)x(t) = 0 \qquad (6)$$

can be either linear or nonlinear, depending, respectively, on whether *a*(*t*) is a fixed function defined by the very system's structure (i.e. given by the producer of the system) or an externally defined *input* of the system.

Since we see in the definition of input an axiomatic point, let us consider this seeming terminology difficulty in detail.

When a fixed time-function that defines the structure of a system arises in a linear system operator, say $a_1(t)$ in the following operator-sum

$$\hat{M}_t = \sum a_p(t) \frac{d^p}{dt^p} \qquad (7)$$

(think e.g. about the equation $\hat{M}_t x(t) = u(t)$ ), then this is, *usually*, a function defined by the producer of the device, and it is the same in any work-state. Such a system is linear. Because of the complete fixation of $a_1(t)$, interpreting the system's action as the map $a_1(t) \to x(t)$ would be unnatural.

If, however, $a_1(t)$ is an input of the system, then such a function has to be seen as *taken from a set of the permitted function*, i.e. it belongs to an available for the (human) operator region in some function space, and not to only one of its points. This is required by both the operational purposes in which the input has to be changed and by the necessity to perform the input scaling test (5), or the additive test of linearity (and thus a linear space is included). Though in each *particular* experiment, the input function is given, we *have* to consider its possible changes, and, in particular, such expressions as $ku(t)$, *with an arbitrary k*. This is the disquieting sense in which the input function is "given". (Letting first $u(t)$ to be an essentially positive, known function describing the state of one's bank account, take then in $ku(t)$, $k$ as 1/2 or -2.)

It becomes clear that though **u** is, in principle, known, [A(**u**)] in (3) is not as "given" as [A($t$)] in (2). (Space with its operations cannot be replaced by a point/function.) If the scaling factor, or the input-function waveform can be changed, then, in fact, the system is not closed, i.e. it is not completely known. This agrees with the classical and most natural view of "input" as a kind of "interface" of a system with the external world, not as an internal fixed characteristic of the system.

However, as a rule, when considering (6), one does not ask what the known function $a(t)$ is, and one's immediate reaction is: "*Everybody knows that this is a linear equation!*". Such a reaction, obviously, follows from one imagining (6) to be the particular case of

$$\frac{dx}{dt} + a(t)x(t) = u(t) \qquad (6a)$$

with the input $u(t)$ taken as zero. *One simply always assumes that the right-hand side of an equation is the reserved place for the "input"*. Thus $a(t)$ in (6a) and (6) is not seen by one as any input.

However one has to remember that:

1. "*Right-hand* side" is not a mathematical, but a physiological concept, and such, -- however traditional, -- a view of the equational structure has no relation to mathematical rigor.

2. Modern electronics systems can have numerous unusual ports, and thus the input function(s) can appear in different equation terms. This is not as in the old mechanical problems of the period when differential equations appeared, and when the outlook on such equation was established.

3. It is insufficient to know that $a(t)$ is given; one has to know what is the role of this function in the real system. No general reason exists why $a(t)$ in (6) can not be an input, and why in (6a) both $u(t)$ and $a(t)$ can not be inputs. Note that it can be, in particular, that $a(t) = u(t)$ in (6a).

Solving (6),

$$x(t) = x(0)e^{-\int_0^t a(\lambda)d\lambda},$$

one sees that the map $a(t) \to x(t)$ is not additive (not linear), but multiplicative. Thus, if $a(t)$ is the input, (6) is a nonlinear equation.

One notices that this is a standard nonlinearity in the sense of our basic equations. Indeed, while seeing in each case $a(t)$ as **u**(t), one can either see (6) as a case of (4),

$$\frac{dx}{dt} = -a(t)x,$$

or, with a minor change in writing,

$$\frac{dx}{dt} = -x a(t),$$

as a case of (1a), in which $[A] = 0$ and $[B(\mathbf{x})] = -x$. See also [6].

Returning to the operator of type (7), we can say that if such an operator includes functions-coefficients that *by themselves* (via some input) influence the functions on which the operator is intended to act, then this operator is nonlinear. Thus, in the proper system context, a mathematical operator formally having a linear form can become nonlinear.

Of course, it does not yet follow from the axiomatic importance of classifying the **u**-systems as nonlinear that any nonlinear effect that can be obtained in **x**-systems can be also obtained in a **u**-system.

Example of a circuit with **u**-nonlinearity is given in [33].

There is another interesting case of nonlinearity to be separately noted. It can occur (see Section 4.3) in a nonlinear switched system that for some values of systems parameters a nonlinear term becomes a known time-function. Namely, the equation

$$(\hat{L}x)(t) + f(x) = k\xi(t) \qquad (8)$$

where $k\xi(t)$ is the input and $f(x)$ is a nonlinear function, can became (because of a saturation of $f(.)$, see Section 4.3) *for some values of the parameters included in the linear operator* $\hat{L}$

$$(\hat{L}x)(t) + \zeta(t) = k\xi(t) \qquad (8a)$$

where $\zeta(t) \equiv f(x(t))$ is a known time-function that relates, just as $f(x)$ in (8), to the system structure, and not to the input. The left-hand side is now an *affine*, i.e. a *nonlinear* form by $x$ (the test of linearity does not pass because of the fixed term), and thus also for this specific range of the parameters of $\hat{L}$ the equation remains nonlinear and no drastic change in the associated physical system occurs. Of course, the issue is delicate because physically the term $\zeta(t)$ is obtained (Section 4.3 for details) only for some nonzero, sufficiently large $k$. Nevertheless, the known function appears *not* as a part of the input.

Though the system behind the example of Section 4.3 is practically important, the function $f(x)$ leading to the affine nonlinearity, is rather specific, and we shall not include the affine case in the main classification below, considering it as part of the **x**-nonlinearity. However, one sees here too how important it is to know/see whether or not a known function belongs to the input.

## 3. Linear and nonlinear switchings

Switching systems, linear and nonlinear, are a very important background for the present study. Following [3], we exclude in these systems any "analytical nonlinearity", leaving the possibility of nonlinearity to arise only because of the switchings.

Considering any switched system, -- either linear or nonlinear, -- as

$$d\mathbf{x}/dt = [A(t,\mathbf{t^*})]\mathbf{x} + [B(t,\mathbf{t^*})]\mathbf{u}(t), \qquad (9)$$

where $\mathbf{t^*}$ is the set of switching points $\{t_k\}$ that define the system operation, i.e. at which the elements are changed, we have the case of linearity (2) if

$$\mathbf{t^*} = \mathbf{t^*}(t), \qquad (10)$$

i.e. if $\mathbf{t^*}$ is defined by known functions (usually external generators), or, simply, is *prescribed*, and the case of nonlinearity (1a) if

$$\mathbf{t^*} = \mathbf{t^*}(\mathbf{x}), \qquad (11)$$

i.e. if $\mathbf{t^*}$ is defined by initially unknown functions that have to be found when solving the system. The case of

$$\mathbf{t^*} = \mathbf{t^*}(\mathbf{u}) \qquad (11a)$$

is also nonlinear.

Thus, for the switched systems, (10) compactly represents (2), (11) represents (1a), and (11a) represents (3).

Respectively to (10), (11) and (12) we speak about "*t*-", "**x**-", and "**u**-" SS.

In the research scheme introduced by [3], for (10) and (11) $\mathbf{t^*}(.)$ is defined by the functions (either known/prescribed, or not) that are inputs of the comparators that determine the level-crossings at which the pulses are generated to trigger the switches. Figure 1 schematically illustrates this.

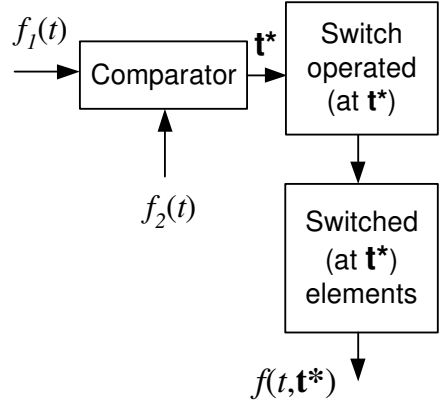

Fig. 1: A switching subsystem somewhere inside the given system. Starting from the intersections of the shapes of $f_1(t)$ and $f_2(t)$, which define **t***, we come to a function $f(t,\mathbf{t}^*)$ measured on the switched unit.

   If both $f_1(t)$ and $f_2(t)$ are prescribed we have a linear (LTV) system, and if at least one of these functions is one of the state-variables to be found, then the system is nonlinear;  **t*** = **t***(**x**).  The case of **t***(**u**) is nonlinear too. (For a **u**-system, it can occur that changing a scaling factor may not influence **t*** for every input waveform; then the generalization of the Small Theorem mentioned in Section 1, associated with considering *additive changes* in **u**, can be important.)
   In order to see how **t*** = $\{t_k\}$ can be included within the system's state-equations, assume that one creates a switching system by replacing a capacitor C in a linear time-invariant (LTI) circuit described using matrix [A(C)], by a switched unit $C_1 \leftrightarrow C_2$. Assume that at some instant $t_1$, known/prescribed, *or not*, $C_1$ is replaced by $C_2$. Considering that the coefficients in Kirchhoff's equations, which define [A] are given as some instantaneous time-functions, one can simply replace in [A(C)] the parameter C by the time-function of the type

$$C(t; t_1,\ldots) = \ldots C_1 u(t_1-t) + C_2\, u(t-t_1) + \ldots , \qquad (12)$$

where u(*t*) is the step function (u(z) = 0 for z<0, and 1 for z>0), obtaining instead of [A(C)] some matrix of the type

$$[A(C_1, C_2, t, \mathbf{t}^*)] , \quad (\mathbf{t}^* \supset t_1) \qquad (13)$$

   According to the said about equation (3), such a switched system may be either LTV or nonlinear, depending on whether or not the functions whose mutual crossings define **t*** are known; i.e. depending on whether or not **t*** is known.
   If the switched system is nonlinear, then one can, perhaps, obtain stabilization of oscillations in it (a limit cycle), or chaos (an attractor).  If the system is linear, then some needed linear response to a port-input can be obtained, and the system can be well controlled, e.g., in the sense of its power consumption.

This is a simple and very general way of classifying switched systems as linear or nonlinear, and is suitable for a designer who actually chooses the functions at the inputs of the comparators. See the circuit examples in [3].

Our final scheme of classification of switched systems is shown in Fig. 2. This is, essentially, also a classification of more general singular systems, e.g. sampling systems, or systems with singular passive elements.

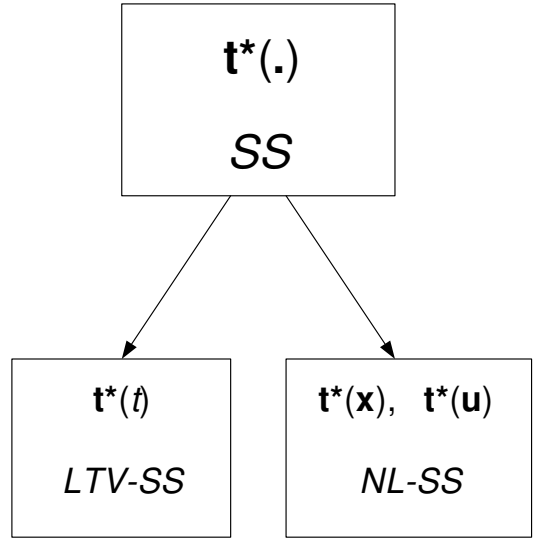

Fig. 2: The general scheme for classification of switched system (SS).

## 4. Switching instants as functionals.

Let us return to Fig. 1. If we take in it $f_1(t)$ as $x_1(t)$ and (for simplicity) $f_2(t)$ as a constant level $D$, then the associated $t_k$ can be written using the inverse function, as $x_1^{-1}(D)$. However the inverse function (certainly existing in the vicinity of a level-crossing of $x_1(t)$), is a local one and it is not a very suitable concept here. In general, the notation **t\***(**x**) just means that **t\*** is *defined* by **x**, and it is not necessary here that some analytical dependence of a $t_k$ on an $x_k$ be given. Indeed, the switching instants are, finally, some numerical values, and, using the terminology of the calculus of variations, the map $\mathbf{x}(t) \rightarrow \mathbf{t^*}$ is of the type of a *functional*, i.e. *a map of a function on a number* (or a vector-function on a set of numbers). The simplest way of defining a $t_k \in \mathbf{t^*}$ is through the concept of level-crossing of a time-function and the use of comparators as illustrated in Fig. 1. For instance, $x(t) = 0$ (i.e. $f_1(t)$ is $x(t)$, and $f_1(t) \equiv 0$) can be the equation for a $t_k$ in the nonlinear case.

In any case, we assume that existence of a map $\mathbf{x}(t) \rightarrow \mathbf{t^*}$ (or $\mathbf{u}(t) \rightarrow \mathbf{t^*}$) should be a general rule for any nonlinear switched circuit.

### 4.1. Two examples using scalar equation form

In [8] the linear-form equation (*a, b, c* -- constants)

$$\frac{d^2x}{dt^2} + a\frac{dx}{dt} + bx(t) = c\sin\omega t \qquad (14)$$

with the "forced" condition of the mirror-type reflection of *x(t)* from the time axes:

$$(dx/dt)(t_k^+) = - (dx/dt)(t_k^-), \qquad t_k: x(t_k) = 0, \qquad (15)$$

is considered. Obviously, $t_k$ are included via $\{t- t_k\}$ in the solution of (8); $x(t) = F(t,\{t- t_k\})$ with a known function $F(.,\{.\})$, and for any certain *p*, the condition $x(t_p) = 0$ becomes $F(t_p;\{t_p - t_k\}) = 0$, i.e. all $t_k$ may be thus found. System (14,15) is *nonlinear* because zeros belonging to the function to be found are used. The computer experiment [8] indeed has shown that for some ranges of the parameters *a*, *b* and *c*, the sequential development of the process leads to a chaotic *x(t)*, which is a feature of a nonlinear system.

Another chaotic equation, this time with a nonlinear singular *characteristic*, which is a more standard case, is described in [9]. In the equation

$$\frac{d^2x}{dt^2} + x(t) = \sin(\sqrt{2}\,t) + s(x) \qquad (16)$$

*s(x)* is the piecewise linear map defined as *max*{-1, 5*x*} if $x < 0$, and as *min*{5*x*,1} if $x > 0$, i.e. the instants of singularity at which an LTI equation is switched are defined by *crossings* by the function 5*x* the levels -1 and 1.

Introducing the unknown "switching" (singularity) instants $\{t_k\} = \{t_k(x)\}$, defined by the relevant level-crossings of *x(t)*, we can write *s(x)* as some $f(\{t-t_k(x)\})$, having

$$\frac{d^2x}{dt^2} + x(t) = \sin(\sqrt{2}\,t) + f(\{t - t_k\}) \qquad (16a)$$

where $f(\{.\}$ is, in principle, known. It is found in the intervals of the type $t_k < t < t_{k+1}$, (note that $t_k(x) = t_k(x)$) which are defined by either (when $s(x) = x$, i.e. |*x*| <1/5) the equation

$$\frac{d^2x}{dt^2} = \sin(\sqrt{2}\,t) \qquad (16b)$$

or (when |*s(x)*| = 1, i.e. |*x*| >1/5) by the equations

$$\frac{d^2x}{dt^2} + x(t) = \pm 1 + \sin(\sqrt{2}\,t). \qquad (16c,d)$$

The **t\*(x)**-nonlinearity is obvious, directly following from the singularity of $s(x)$, even though it is not easy here to write an intermediate constructive solution of (16) in terms of $\{t_k\}$. However, in the example of Section 4.3 such a constructive solution is not difficult.

### *4.2. Switching "between linear systems"*

Assume that all of the elements of [A] that are changed, are changed at the same instants. Of course, one can assume that when only some (e.g. one) of the elements of [A] are changed, all the other elements undergo zero changes (i.e. switching to the same value) *at the same instants*.

Then, such functions as $u(t_1-t)$ and $u(t-t_1)$, appearing in the matrix elements of type (12), can be taken from the matrix, and (12) is generalized to

$$[A(t,t_k)] \;=\; [A]_1\, u(t_k-t) \;+\; [A]_2\, u(t-t_k) \;+\; \ldots \;, \qquad (17)$$

where $[A]_1, [A]_2, \ldots$ are some fixed (because the switching is between LTI elements) numerical matrices.

Assume now that all the matrix-coefficients in (17) are taken (repeated in some way) from the set composed of only two fixed matrices, $[A]_a$ and $[A]_b$, i.e.

$$[A(t,t_k)] \;=\; [A]_a\, u(t_k-t) \;+\; [A]_b\, u(t-t_k) \qquad (18)$$

which is $[A]_a \rightarrow [A]_b$, and then back ($[A]_b \rightarrow [A]_a$):

$$[A(t,t_{k+1})] \;=\; [A]_b\, u(t_{k+1}-t) \;+\; [A]_a\, u(t-t_{k+1}) \qquad (18a)$$

(say, (18) for all-even, and (18a) for all-odd indices). That is, [A] alternates the two "states", $[A]_a$ and $[A]_b$, according to some specific criteria for $t_k(.)$.

In such a case, instead of speaking about the switching of elements in a certain LTV or nonlinear system, one can speak about switching "*between* two LTI systems", each corresponding to one of the two given numerical matrices, $[A]_a$ or $[A]_b$.

This kind of terminology, though without writing (17) or (18), i.e. not explicitly introducing **t\***(.) in the equations, but *indeed creating two complete LTI subsystems and switching from one to another*, is used in work [10] devoted to the generation of a chaotic process. **t\*** = **t\*(x)** in [10]. However this is obtained not via the level-crossings of some of the $x_k(t)$ but via some constraints on the norms of **x**(t) which require the calculation of some associated measures.

Unfortunately, in stressing the linearity of the two LTI subsystems, work [10] does not mention the nonlinearity of the *whole* system. However, from our positions, it is immediately seen from $\{t_k\} = \{t_k(\mathbf{x})\}$ that the system is nonlinear.

Thus, regarding such works as [8-10] our classification of the nonlinearity of switched systems seems to be heuristically useful. We continue, however, with other important examples.

### 4.3. An integral equation related to the case of a "physical switching" inside a singular passive element

Usually, determination (calculation) of $\{t_k\}$ is done only at the very final stage of solving a problem where the map **x → t*** takes place. A relevant example is given in [12-14]; it can be formulated in terms of the differential electrical circuit equation,

$$L\frac{di}{dt} + A\,\text{sign}[i(t)] + \frac{1}{C}\int i(t)\,dt = f(t), \qquad (19)$$

where $f(t)$ is a $T$-periodic given function, which is equivalent to the integral equation of the type (using more common mathematical notations)

$$x(t) = \varphi(t) - \mu \int_{-\infty}^{t} h(t-\lambda)\,\text{sign}[x(\lambda)]\,d\lambda \qquad (20)$$

where $h(t)$ is the current shock response (Green's function) of the oscillatory subcircuit to which the parameters L and C of (19) relate, and $\varphi(t) = (h*f)(t)$ is a known $T$-periodic function having zero average, whose zerocrossing features are important.

This equation plays an important role in a simple version of a nonlinear theory of fluorescent lamp circuits. See [12-14] and the works quoted there.

It can be proved that for $\mu$ small enough $x(t)$ in (20) is a $T$-periodic zerocrossing function, having the same density of the zero-crossings as $\varphi(t)$. Using Fourier series for the time-function $\zeta(t) \equiv \text{sign}[x(t)]$, one obtains from (20) for such $\mu$

$$x(t) = \varphi(t) - \mu\zeta(\{t-t_k\}) \qquad (21)$$

with known function $\zeta(\{.\})$, i.e.

$$x(t) = F(t,\{t_k\}) \qquad (21a)$$

with $F(.,\{.\})$ known, but $t_k$ themselves still unknown.

One observes from (21) that the zerocrossing features of $x(t)$ require [14,13] that $d\zeta/dt$ be limited.

Starting from $\mu = 0$, and increasing $\mu$, we observe, in the general case, "movement" of the zero-crossings of $x(t)$ from their initial positions defined by $\varphi(t)$; the range for this "movement" is defined by appearance of some touching of the time-axes by $x(t)$. There is an interesting case [5,13,14] when $t_k$ are *unmoved* (constant) within these bounds for $\mu$. In this case, $x(t)$ given by (21a) is already completely known. However, in the general case, (21a), is just a stage of the solution.

The problem of solving (19) is thus reduced to the algebraic problem of determination of $t_k$. The equations of the type

$$x(t_k) = 0 \qquad (22)$$

which *define* $t_k$, become the *constructive (usually transcendent) equations* of the type

$$F(t_k, \{t_p\}) = 0, \qquad (23)$$

and all such equations together define all of the $t_k$. In view of (21), (23) is

$$\varphi(t_k) = \mu \zeta(\{t_k - t_p\}), \quad \forall k, \qquad (23a)$$

while for the mentioned case of $t_k$ "unmoved", both sides of the latter equalities are zero. Equation (23) is an example showing how the map **x**→ **t*** can be realized.

This formal simplicity of the solution of (20) in terms of the zero-crossings even can create the impression that (20) is not an integral equation at all, rather a set of some algebraic equations. This seems to be paradoxical since we started from differential equation (19). However for the investigation of the possibility of $x(t)$ having the zerocrossing features of $\varphi(t)$, the integral features of (20), mainly the boundedness of $d\zeta/dt$ [14], are important. In other words, *the dynamics of the system is expressed in* (*23a*) *in terms of the structural stability of the zerocrossing features of x(t)*.

Finally, the case when $A\text{sign}[i(t)]$ in (19) becomes a known function that does not, however, belong to the input, the nonlinearity of the equation is kept by the affine form of the left-hand side; certainly, the physical system is also not changed essentially in this case; the fluorescent lamp remains on its place.

Among other relevant examples, the use of **t*(x)**-nonlinearity to stabilize the amplitude of parametric oscillations can be mentioned; see [12] and references given there.

## 5. Some physical systems

We turn now to some physical systems where the suggestion to see the nonlinearity as an influence of **x** on the structure, and the concept of the **t*(x)**-nonlinearity are useful. Hopefully, such arguments can be helpful in making system theory principles a part of one's general education.

We consider first an analytical "**x**-system", and in Section 5.2 a **t*(x)**-system.

### *5.1. Linearization of a nonlinear system; modeling of a physical system whose very structure is associated with the "input-output" transferal of a signal, i.e. is dependent on "x".*

The suggestions of seeing in (1a) a direct connection of the structure of the system with its nonlinearity, and the possible transfer to a linearized system of type (24), where followed for a problem of *vorticed liquid flow* in [20]. Starting from an analogy to $d\mathbf{x}/dt = [A(\mathbf{x}(t))]\mathbf{x}$, we shall pass on to an equation of the type

$$d\mathbf{x}/dt = [A(\mathbf{x}_o(t))]\mathbf{x}, \qquad (24)$$

i.e. the autonomous form of (2).

Even when not knowing anything about the Navier-Stokes equation, but observing the flow, one understands in view of (1a) that *since the velocity vector field is both the solution vector-function and the structure of the liquid "system", this system is* *nonlinear*. Thus, for instance, possible chaotic movement (i.e. turbulence) can be understood as a kind of chaos obtained in a nonlinear system.

Assume that a perturbation of a flow (that initially may be, for simplicity, laminar) is made at a point in the stream. This perturbation is expressed in some vortices introduced in to the flow, which are carried by the flow as an input "signal", and, at the same time, are also a part of the flow. *Thus, following the line of thought dictated by (1a), we see the "input-output signal" as a part of the very structure of the "system".*

Since it is clear that the "signal" propagating with the flow is being somewhat spread and changed, it is also clear that the transfer function of the standard type, which would express transfer of the "signal" *as it is*, i.e. only with a *time delay* $t_x$ (the time needed for the perturbation to come to the point of observation $x$, taken in the direction of the flow),

$$H_o(s) \sim \exp\{-t_x s\}, \qquad (25)$$

is non-realizable. The work-hypothesis of [20] is that if the perturbation is changed weakly, the true $H(s)$ has, however, to be close to $H_o(s)$. Using

$$H(s) = (1+st_x/n)^{-n}, \qquad (26)$$

(but, of course, not $(1-st_x/n)^n$ which enhances, as one sees taking $s \to \infty$, high frequencies and requires negative viscosity in (27)) as an approximation to (25), [20] shows that this system function agrees with a linearized Navier-Stokes equation, while $n$ is the simultaneously obtained Reynolds number in which $x$ is the "parameter" of the distance.

Consider the Navier-Stokes equation [20-22] in which, following [20], we omit the term with the gradient of pressure:

$$\frac{\partial \vec{v}}{\partial t} + (\vec{v}\nabla)\vec{v} = \nu \Delta \vec{v} \qquad (27)$$

where $\vec{v}$ is the velocity vector field (i.e. $v_x$, $v_y$, and $v_z$, are our state-variables) and $\nu$ is the kinematic (divided by the density of the liquid) viscosity. In the 1D-case (27) becomes the known Burger's equation $u_t + u u_x = a u_{xx}$ used for demonstrating the specificity of some hydro- and aero-dynamic phenomena. The ignorance of the pressure ($p$) is, of course, unacceptable for general physical analysis (the complete dynamic vector equation for incompressible liquid, $\frac{\partial \vec{v}}{\partial t} + (\vec{v}\nabla)\vec{v} = \nu \Delta \vec{v} + \frac{1}{\rho}\nabla p$ includes four unknowns, and the condition for incompressibility, $\nabla \vec{v} = 0$, has to be added), but $v_x$, $v_y$, and $v_z$ are the parameters of $\mathbf{x} = \{\vec{v}, p\}$ which relate to the immediately seen "structure" of the liquid system, and thus our comparison below of (28) with (1a) is legitimized, even though $[A_\Sigma(\vec{v})]$ below is not the full physical $[A(\mathbf{x})]$.

Equation (27) can be easily rewritten in the spirit of (1a), as

$$\frac{\partial \vec{v}}{\partial t} = -[A(\vec{v})]\vec{v} + [\alpha]\vec{v} \qquad (28)$$

where the nonlinear matrix-operator

$$[A(\vec{v})] = \begin{pmatrix} \vec{v}\nabla & 0 & 0 \\ 0 & \vec{v}\nabla & 0 \\ 0 & 0 & \vec{v}\nabla \end{pmatrix} = [I](\vec{v}\nabla)$$

includes the unit matrix $[I]$, and the scalar differential operator

$$\vec{v}\nabla \equiv v_x \frac{\partial}{\partial x} + v_y \frac{\partial}{\partial y} + v_z \frac{\partial}{\partial z} \quad ,$$

and

$$[\alpha] = \nu[I]\Delta \equiv \nu[I](\frac{\partial^2}{\partial x^2} + \frac{\partial^2}{\partial y^2} + \frac{\partial^2}{\partial z^2})$$

is a linear matrix operator.

Introducing

$$[A_\Sigma(\vec{v})] = -[A(\vec{v})] + [\alpha] = [I](-\vec{v}\nabla + \nu\Delta) = \begin{pmatrix} -\vec{v}\nabla + \nu\Delta & 0 & 0 \\ 0 & -\vec{v}\nabla + \nu\Delta & 0 \\ 0 & 0 & -\vec{v}\nabla + \nu\Delta \end{pmatrix},$$

we can rewrite (28) as

$$\frac{\partial \vec{v}}{\partial t} = [A_\Sigma(\vec{v})]\vec{v} \ . \qquad (28a)$$

Matrix $[A_\Sigma]$ represents the system's "structure" and is of the type $[A(\mathbf{x})]$ in (1a).

The linearization of (27) means in [20]

$$\frac{\partial \vec{v}}{\partial t} + (\vec{v}_o\nabla)\vec{v} = \nu\Delta\vec{v} \qquad (29)$$

where $\vec{v}_o$ is the *given average velocity of the flow*. In state-space terms, this linearization means the replacement in (28) of $[A(\vec{v})]$ by $[A(\vec{v}_o)]$, or in (28a) $[A_\Sigma(\vec{v})]$ by $[A_\Sigma(\vec{v}_o)]$:

$$\frac{\partial \vec{v}}{\partial t} = -[A(\vec{v}_o)]\vec{v} + [\alpha]\vec{v} = [A_\Sigma(\vec{v}_o)]\vec{v} \ ,$$

which is similar to (24).

Orienting the x-axis in the direction of $\vec{v}_o$, taking into account that for the "inertial part" of the perturbation of $\vec{v}$ (i.e. the purely delayed "input signal"), denoted as $\vec{v}_i$, the left-hand side of (29) is precisely zero, and writing

$$\vec{v} = \vec{v}_o + \vec{v}_i(x - v_o t) + \varepsilon(x,t),$$

where $\vec{\varepsilon}(x,t)$ is the *small* deviation occurring in the introduced perturbation $\vec{v}_i$ after it has passed the distance $x$, [20] derives from (29) the following equation for $\varepsilon(x,t)$:

$$\frac{\partial \vec{\varepsilon}}{\partial t} + v_o \frac{\partial \vec{\varepsilon}(x,t)}{\partial x} = \nu \frac{\partial^2 \vec{v}_i(x - v_o t)}{\partial x^2} \quad . \tag{30}$$

The Laplace transform of (30) by $t$ is then done, turning this PDE to a simply solved ODE with derivation by only $x$.

The assumption of the transfer function (26) is finally shown in [20] to be reasonable, in agreement with the linearization of the Navier-Stokes equation, and '$n$' appears to be an analogy of Reynold's number, including $x$ as the distance parameter.

As the matter of fact, -- it was this research which originally led the author to the opinion that in some problems form (1a) may be generally better than (1). However, the approach to **x**-systems via **t\***-systems also has some certain advantages, and we continue with *singular* systems.

### *5.2. An ensemble of colliding balls: "thermalization" as a result of the t\*(x) nonlinearity*

It may be observed that for an ensemble of many colliding particles, the chaotic movement of the particles is obtained because of the nonlinearity of the "switching" type. Indeed, the instants when the strikes (collisions) of the particles appear depend on the trajectories $r_i(t)$ of the particles, which are our state-variables, i.e. this is a **t\*(x)**-nonlinearity. This consideration even leads us to the general assumption that:

*Any ensemble of colliding particles where chaotic distribution of the movement is obtained is a nonlinear system with a kind of nonlinearity which is close to that found in nonlinear switched systems.*

Of course, chaos appears here as a possible *indication* of nonlinearity whose existence should be shown directly.

If the above assumption is correct, then, in particular, the tendency to thermodynamic equilibrium ("thermalization"), occurring in many systems, is a nonlinear **t\*(x)**-process.

In order to better see the **t\*(x)**-nonlinearity of the ensemble of colliding particles, let us consider the particles as small rigid balls whose collisions are momentary.

The spatial position $r_i(t)$ of a ball number '$i$' is defined by its initial position and the $\delta(t)$-type forces that arise in the collisions of the ball with other balls at the instants $t_{i,j}$ that are roots of equations of the type $r_i(t) - r_j(t) = 0$, $j \neq i$. These forces may be written as $F_{ij} = F_o(v_i(t_{ij}^-), v_j(t_{ij}^-)) \delta(t - t_{ij})$, where $v_i = dr_i/dt$, i.e. $v_i(t_{ij}^-)$ and $v_j(t_{ij}^-)$ are the velocities of the balls just before the collision. Since the function $F_o(.,.)$ can be found from energy and impulse conservation laws, its form is universal for the collisions, and let us focus only on the time-dependent factor $\delta(t - t_{ij})$ that includes the shift.

In view of the dynamic law

$$\frac{d\vec{v_i}}{dt} = \frac{1}{m} \sum_{j \neq i} \vec{F}_{ij} =$$

$$= \frac{1}{m} \sum_{j \neq i} \vec{F}_o(v_i(t_{ij}^-), v_j(t_{ij}^-)) \delta(t-t_{ij}), \quad \forall i,$$

{$v_i(t)$}, and thus {$r_i(t)$}, can be presented as some explicit functions of {$t$-$t_{ij}$}. However since {$t_{ij}$} themselves are defined by the unknown {$r_i(t)$} (our **x** here), the above system of equations is nonlinear in the sense of the shifting (**t\*(x)**-) nonlinearity, i.e. the {$t_{ij}$} are quite as some {$t_k$} in a electronic switched circuit.

We thus can see such an ensemble of the colliding particles as a **t\*(x)**-system.

Wishing to see the reason for the nonlinearity, we just described the basic map {$r_i$}→{$t_{ij}$}. Of course, in order to find the numerical values of {$t_{ij}$}, $\vec{F}_o(.,.)$ has to be determined, and one has to decide whether the collisions are elastic or not, etc..

We conclude that it is possible that the **t\*(x)**-nonlinearity is always around us and is "as old as this world", though, of course, definition of a "system" should be properly extended.

### *5.3. The Gaussian distribution as another aspect of the tendency of a system to a statistically equilibrium state through the nonlinear* **t\*(x)**-*dynamics*

This kind of nonlinearity also causes Gaussian distribution, because this statistical distribution is associated with the tendency of the entropy of the system towards maximum, which is seen as follows.

From Boltzman's formula connecting entropy with probability, $S = k \ln P$, we have $P = \exp\{S/k\}$, and since near its maximum that takes place at some $x = x_o$ ($x$ is a parameter being observed), $S(x) \approx S(x_o) - d(x - x_o)^2$, with a constant $d > 0$, we obtain $P(x) = K \exp\{-(d/k)(x - x_o)^2\}$, where $K = \exp\{S(x_o)^2/k\}$, i.e. near the maximum of entropy of the system in which $x$ is observed, $P(x)$ becomes the Gaussian distribution.

Thus, the coordinates of the movement of a Brownian particle, found inside ensemble of small particles that are in thermodynamic equilibrium, have the Gaussian distribution.

The assumption can be expressed that if the distribution of fish in sea were to be homogeneous in the proper scale, then the movement of the hunting albatross in the air over the sea ([23,24] and the works quoted there) can also be expected to be Gaussian. Seeing/attacking a fish defines the instant $t_k$ of the singularity of the trajectory.

That the movement of the hunting albatross is defined/influenced by the statistically characteristics of the movements in the ensemble of the fishes is obvious. Both the Brownian particle and the albatross can be interpreted as some devices or sensors measuring the statistical parameters of the associated medium through the (event-defined, i.e. "functional" in the sense of Section 4) **t\*(x)**-dynamics of their movement.

This further stresses the role of the **t\*(x)**-nonlinearity in the natural processes, -- a pedagogical/educational point, not to be missed bay a teacher.

We turn now to another important class of singular systems, sampling systems.

## 6. Comments on nonlinear sampling

What was said about **t\***(.) in the cases of switched systems and systems including passive elements with singular characteristics, can be also said about any singular systems, among which *sampling systems* are very important.

For a *linear sampling procedure*, in the functional

$$f(.) \to f(t^*), \qquad (31)$$

$t^*$ is fixed, independent of $f$, and the relevant realization, block-scheme, [5] does not differ strongly from the scheme in Fig. 1. It is correct for (31) that

$$(f_1+f_2)(t^*) = f_1(t^*) + f_2(t^*), \qquad (32)$$

i.e. summing the shapes of the functions and then sampling is the same as to sampling the shapes separately and then summing the results.

Nonlinear sampling appears when the sampling instants are dependent on the input function of the sampling (sub)system, $t^* = t^*(f)$. This thesis (considered also in [5]) is developed in the following subsections and Section 7. The point here is that such nonlinearity avoids prescription of the sampling rate and makes the sampling adaptive in a sense that helps in the analysis of the input function.

### *6.1. The sampling "by definition"*

In the following argument, the input function of the sampling subsystem will be sampled *by itself* at its own level-crossings, which introduces the nonlinearity of the sampling in the simplest possible manner. We just observe that for the nonlinear expressions,

$$(f_1+f_2)(\text{at its } t^*) \qquad (33)$$

is, generally, not equal to

$$f_1(\text{at its } t^*) + f_2(\text{at its } t^*), \qquad (34)$$

and the distinction can be very significant.

Let us introduce the dependence of the sampling instants $\{t_k\}$ = **t\*** on the sampling function $f$, $t_k(f)$, by using level-crossings of $f(t)$ with some fixed constant level $f_{ref}(t) \equiv D$. That is, we shall sample $f(t)$ just at the instants where these level-crossings occur. We shall obviously obtain the sampled values as

$$f(t_k) = D. \qquad (35)$$

If $D$ is *unknown*, it can thus be measured, but, in general, (35) is correct "by definition". Such an equality, illustrated by Fig. 3, ceases, however, to be trivial in the construction of Lebesgue's integral in Section 6.2 when $t_k$ have to be found.

Since different types of nonlinearity can cause chaos, this realizing scheme can be investigated as regards the stability of its operation, and the filtration of noise in $f(t)$.

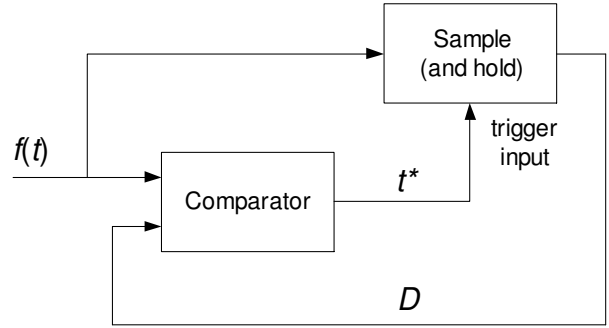

Fig. 3: Sampling "by definition". The instant $t^*$ is defined by the crossing by $f(t)$ of the level $D$, i.e. as $f^{-1}(D)$, and $D$ is simultaneously defined as $f(t^*)$. Mathematically, this is the identity $D = f(f^{-1}(D))$, though not a universal one since $f^{-1}(.)$ can be only *locally* defined, but realizing this scheme and studying its stability can be relevant to modeling the finite sums approximating Lebesgue's integral, considered in Section 6.2.

For *thus respectively sampled* values, the quantity (33) is, generally, not equal to (34). For instance, for $f_1(t) = \alpha t$, and $f_2(t) = \beta t$, where $\text{sign}[\alpha] = \text{sign}[\beta] = \text{sign}[D]$ and thus the level-crossings obviously exist, we obtain

$$(f_1+f_2)(\text{at its } t^*) = D$$

$$(f_1)(\text{at its } t^*) = D,$$

$$(f_2)(\text{at its } t^*) = D,$$

and (33) is $D$, while (34) $2D$. The only exceptional case when (33) equals (34) here is that of $D = 0$.

The situation regarding an oscillating or just a bounded function, when the existence of the level-crossing is not ensured, is somewhat more complicated. For instance, for

$$f_1(t) = A \sin \omega t, \quad f_2(t) = B \sin \omega t, \quad 0 < A < B,$$

there are different cases re the equality of (33) and (34), associated with the possibilities of $D \in (A,B)$ and $D \notin (A,B)$. For $A < D < B$, we have both (33) and (34) equal $D$, but for $B < D < A+B$ (33) is $D$, but (34) is 0.

We conclude that a $t^*(f)$-nonlinearity of the sampling can make (33) strongly different from (34).

### 6.2. *A comment on Lebesgue's integral: nonlinearity of the approximating sums*

That we were taking *zero* in the above comparison of (33) and (34) (as in the case of $D = 0$) *when there is no level-crossings*, has to be justified. This is a decision taken by the analogy with the *measure* that is used for the creation of Lebesgue's integral [15]. In this integral, instead of starting from some portions of the argument, as in Riemann's integral, we start (Fig. 4) from some portions $(\Delta f)_k$ of the function, and seek, at a level $D = f_k$, for associated support on the axis of the argument ($t$), obtaining, in the finite interval of the integration, a set of the associated (supporting) values $(\Delta t)_{k,m}$. Figure 4 illustrates our notations; the level $D = f_k$ contributes several such "blocks" as $D(\Delta t)_{k,1}$ to the approximating sum. It is relevant for us that $(\Delta t)_{k,m}$ are dependent on $f$, and thus the blocks $f_k \cdot (\Delta t)_{k,m}$ are nonlinear by $f$, i.e. the whole approximating sum is nonlinear.

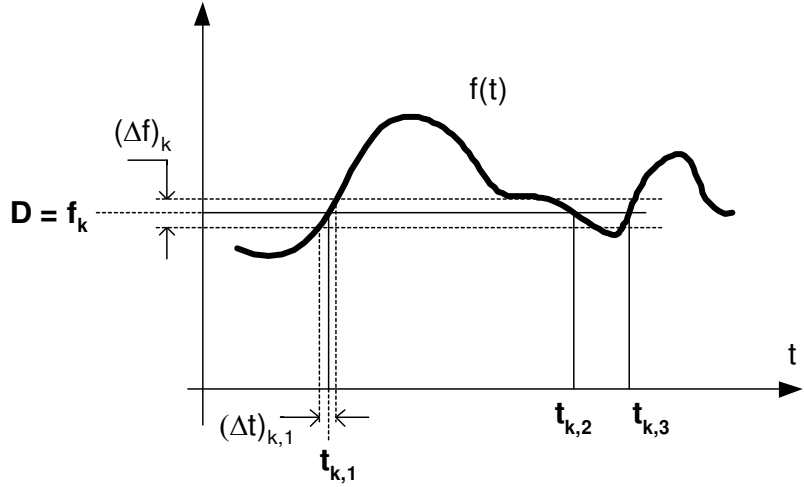

Fig. 4. Construction of the Lebesgue's integral. The intervals $(\Delta t)_{k,m}$ are taken *not* arbitrarily as in Riemann's integral; they depend on $f$. This leads to nonlinearity of the approximating sum.

In simple details, since for the function possessing the needed derivatives

$$(\Delta f)_k = \frac{df}{dt}(t_k)(\Delta t)_k + \frac{1}{2}\frac{d^2 f}{dt^2}(t_k)(\Delta t)_k^2 + \dots ,$$

if $(df/dt)(t_k)$ is nonzero, we have

$$(\Delta t)_{k,m} = \frac{(\Delta f)_k}{(df/dt)(t_{k,m})} , \text{ at the level } f_k = D = f(t_{k,m}) , \forall m. \quad (36)$$

If $(df/dt)(t_k)$ is zero, but the second derivative at $t_k$ is nonzero, then, taking positive values,

$$(\Delta t)_{k,m} = \sqrt{\frac{2(\Delta f)_k}{(d^2 f/dt^2)(t_{k,m})}} , \quad (37)$$

etc., with the value of the derivative taken at each time according to the level $f_k = D = f(t_{k,m})$, i.e. having each time each $(\Delta t)_{k,m}$ dependent on $f$.

The approximating "blocked" sum uses the *measures*

$$meas(f_k, \Delta f_k) = |(\Delta t)_{k,1}| + |(\Delta t)_{k,2}| + \ldots, \text{ for each level } D = f_k,$$

and it is

$$\sum_k f_k \, meas(f_k, \Delta f_k) = \sum_k f_k \left( \sum_m |\Delta t_{k,m}| \right), \qquad (38)$$

having each term nonlinear by *f*, and thus being nonlinear as a whole. Contrary to that, the finite sums used in constructing Riemann's integral of *f(t)* are always linear by *f(t)*.

The nonlinearity of each approximating sum is because of the mapping, via *f(t)*, a finite *given/fixed* set $\{f_k\}$ (or $\{D_k\}$) on the *t*-axis, which creates/chooses the set $\{t_k\}$. This nonlinearity even can be demonstrated by using only *one* point (*D*) on the *f*-axis, having the approximation sum composed of only two blocks.

All of the $\Delta t$ are obtained as equal only if $|df/dt|$ = constant, which requires *f*(.) to be a straight line or some saw-tooth wave.

That Lebesgue's integral exists for a much wider class of functions than Riemann's integral is not immediately important for this point, and while thinking about an application of the nonlinearity, one can consider only easily realizable functions for which both integrals exist. However, an attempt to connect the fact that phase modulation is much better than amplitude modulation in the sense of rejecting noise, with the fact that Lebesgue's integral exists for a much wider class of (not smooth) functions might be interesting. We shall touch here (Section 7) the topic of spectrum analysis, but in a much narrower scope.

### *6.3. Remarks*

Returning to the problem of comparison of (33) and (34) in Section 6.1, we note that the terms in (38) $f_k \cdot (\Delta t)_{k,m}$ are zero if $D = 0$, and/or if there are no crossings, since $D > f_{max}$, geometrically means that all the $(\Delta t)_{k,m}$ became zero. Thus it was accepted regarding (33) and (34) that for bounded $f_1$ and $f_2$, the case of $D > max\{max\{f_1(t)\}, max\{f_2(t)\}, max\{(f_1+f_2)(t)\}\}$, without any zero-crossings, is equivalent to the case of $D = 0$.

One sees from the above why Lebesgue's integral is used more in real, and not complex analysis, -- the nonlinearity associated with the level-crossings is of an essentially real type; the concepts of '>' and '<' characterizing the work of a comparator as in Fig. 1, do not relate to complex numbers. Work [5] connects this fact with the wide use of the *real-valued* δ-function in physics and engineering.

The nonlinearity of the sum in the case of Lebegue's integral and the absence of the prescribed time or frequency unit of measurement for the Lesbegue's integral are two closely connected distinctions between Lebesgue's and Riemann's schemes. In the construction of Riemann's integral, one "chops" *the time-axis sufficiently* finely, obviously adjusting the time unit of measurement to the finest variations of the waveform in order for the sum of the blocked area to be sufficiently close to the precise area/integral. Thus, this construction *includes measurement* of the upper bound of the spectrum of the integrand, by a proper "clock". For the construction of Lebesgue's integral the relevant points on the time (argument) axis are chosen automatically, and no time-unit is prescribed.

This aspect is interesting in view of the method of spectrum analysis of Section 7, where we sample a specific transform of the input function (signal, process) at the

level- (zero-) crossings of the input function, and thus study the frequency features of the input function *without using some a priori given frequency unit, or time measure*, which is a deep distinction, with respect to the usual methods of spectrum analysis. This method inevitably must avoid *Riemann integration* of the instantaneous power of the signal. One notes, however, that the deletion of the unit of measurement would be also obtained if one could use Lebesgue's integration of the power in the hardware.

## 7. The $\psi$-transform and the nonlinear sampling

Consider the following *nonlinear and singular transform* $f \rightarrow \psi$ of the input function $f(t)$ ($\langle f \rangle = 0$)

$$\psi(t) = \int_0^t f^2(\lambda) sign[f(\lambda)] d\lambda . \qquad (39)$$

The point is that $\psi(t)$ may be assumed to be limited, at least for the period of the processing, and thus can be generated in an analog (quick) manner. *Numerical integration of $f^2(t)$* which leads to the *unlimited* (and thus irrelevant for analog integration) "energy" function

$$E(t) = \int_0^t f^2(\lambda) d\lambda \qquad (40)$$

(which is a "competitor" of $\psi(t)$ in the procedure below) is rejected because it requires prescribing rate of sampling of $f^2$, i.e. giving a time or frequency unit that we wish to generally avoid here.

Consider *sampling* of $\psi(t)$ by a sampling system *at the zero-crossings* $\{t_k\}$ *of $f(t)$*. Contrary to the sampling done by means of a prescribed generator of sampling pulses, sampling at zero-crossings of an input function is a nonlinear procedure of the **t\*(u)**-type, or, rather, of **t\*(x)**-type, since $f$ is unknown. Thus, the nonlinearity of the problem is not only because the map $f \rightarrow \psi$ is nonlinear; the nonlinearity of the **t\*(x)**-type is necessary for the estimation of the spectrum of $f(t)$ to be done without any clock or band-pass filters.

Figure 5 schematically illustrates the sampling. That thus sampled values of $\psi(t)$ are its *extreme* values is very easy to see ([17] or [18]).

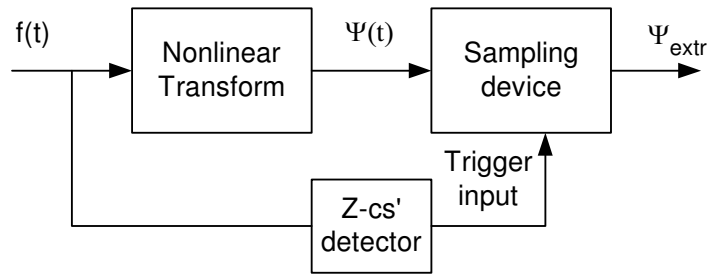

Fig. 5. The schematic for obtaining $\{\psi(t_k)\}$, these parameters to be used for estimation of the average period/frequency of the process $f(t)$. The method thus includes the nonlinearity of **t\*(x)**-type, or **t\*(u)**-type, depending on the outlook on $f(t)$. This scheme is relevant/effective in the case when the integration inside block "Nonlinear Transform" is Riemannian. If the integration is Lebesgue's one, then, as is seen from the basic formulae of the method, one can obtain the estimation of the average period, also not using any prescribed time-unit, but not employing the zerocrossings.

Using the obtained sequence $\psi(t_k)$, we create the following arithmetic average, measured on line,

$$\frac{\sum_{1}^{N}(-1)^k \psi(t_k)}{N},$$

where $N$ is the number of the registered input zero-crossings. (Below $N$ will completely replace '$t$'.) We also use the online-measured "average power"

$$P = \frac{1}{t}\int_0^t f^2(\lambda)d\lambda = E(t)/t \qquad (41)$$

of the signal, obtained on line by *the low-pass filtering* of $f^2(t)$. These two values lead us to the possibility of estimating the average period (frequency) of a compact spectrum of the signal, which is the basic step (to be then repeated in some new channels of the processing system [18]) in the in estimation of the spectrum.

Assuming that on average there should be two zero-crossings per *average period* $T_a$, we define $T_a$ according to the equality

$$N \approx 2\frac{t}{T_a}, \qquad (42)$$

i.e. as

$$T_a \approx 2\frac{t}{N} = 2\frac{E(t)}{NP}. \qquad (43)$$

Of course,

$$NT_a < t < (N+1)T_a, \quad N >> 1,$$

and intending to make $N$ be the main parameter, we shall denote $t$ as $t_N$.

We perform the following transformation of $E(t)$ [17] in which the main part/term is associated with the integral over the interval $(t_1, t_N)$:

$$E(t) = \int_0^t f^2(\lambda)d\lambda = \int_0^{t_1} f^2(t)dt + \int_{t_1}^{t_N} f^2(t)dt + \int_{t_N}^t f^2(\lambda)d\lambda$$

$$= \int_{t_1}^{t_2} f^2(t)dt + \int_{t_2}^{t_3} f^2(t)dt + \int_{t_3}^{t_4} f^2(t)dt + ... + \int_0^{t_1} f^2(t)dt + \int_{t_N}^t f^2(t)dt$$

$$= \int_{t_1}^{t_2} f^2(t)\,\text{sign}\, f\, dt - \int_{t_2}^{t_3} f^2(t)\,\text{sign}\, f\, dt + \int_{t_3}^{t_4} f^2(t)\,\text{sign}\, f\, dt - + ...$$

$$= [\psi(t_2) - \psi(t_1)] - [\psi(t_3) - \psi(t_2)] + [\psi(t_4) - \psi(t_3)] - + ...$$

$$= -\psi(t_1) + 2\psi(t_2) - 2\psi(t_3) + 2\psi(t_4) + ...$$

$$= \psi(t_1) - \psi(t_N) + 2\sum_1^{N/2} \left[\psi(t_{2p}) - \psi(t_{2p-1})\right] + \int_0^{t_1} f^2(t)dt + \int_{t_N}^t f^2(\lambda)d\lambda.$$

(44)

Since as $t \to \infty$, $E(t) \to \infty$, for $N$ large the terms in (44) which are not bracketed are relatively small compared to the value of the sum, and we can write, for the large $N$, (44) as the sum of the (positive) bracketed values:

$$E(t) \approx 2\sum_1^{N/2} \left[\psi(t_{2p}) - \psi(t_{2p-1})\right].$$

This yields according to (40)

$$T_a = \lim_{t \to \infty} \frac{2E(t)}{NP(t)} =$$

(45)

$$= \lim_{t \to \infty} \frac{4\sum_1^{N/2}\left[\psi(t_{2p}) - \psi(t_{2p-1})\right]}{N(t)P(t)}.$$

In this expression we consider the average $\frac{\sum[\cdot]}{N}$ as a limited function of the generally unlimited variable $N$.

We finally write (45) in terms of only the countable zero-crossings

$$T_a = \lim_{N \to \infty} \frac{4\sum_{1}^{N/2}\left[\psi(t_{2p}) - \psi(t_{2p-1})\right]}{NP(t_N)} . \qquad (46)$$

The central parameter finally is *N* and not *t*.

It may be assumed that

$$P = \lim_{N \to \infty} P(t_N) = \lim_{N \to \infty} \frac{1}{t_N}\int_0^{t_N} f^2(\lambda)d\lambda$$

exists, but in any case, we can assume $P(t_N)$ in (46) be limited.

For a *T*-periodic *f(t)*, such that $f(t + T/2) = -f(t)$, this procedure is very simply graphically illustrated in [16], and one sees that also for a non-periodic process it can be assumed that *on average*, $sign[\psi(t_{2p-1})] = - sign[\psi(t_{2p})]$, i.e. (46) can be simplified (notice the change in the upper bound of the sum) to

$$T_a = \lim_{N \to \infty} \frac{4\sum_{1}^{N}|\psi(t_k)|}{NP(t_N)} . \qquad (47)$$

In the simple periodic case of [16], we obtain from by taking $N = 2$ that

$$T = \frac{4\psi(t_1)}{P}$$

($\psi(t_1) = -\psi(t_2) > 0$), which is also obtained from (47) rewritten as

$$T_a = \frac{4\langle|\psi(t_k)|\rangle_N}{P(t_N)} \qquad (47a)$$

with the averaging ($\langle\rangle_N$) by *N*.

No clock or unit of time or frequency is used. It is just required that some limitation the spectrum of *f(t)* to be *roughly* known, in order the integrator, the sampler, and other involved devices to work. The latter is however, a general requirement of any technical device, and this is very far from giving any unit of measurement.

This is the basic step of the procedure of spectrum analysis based on the sampling of the $\psi$-transform at the zero-crossings of *f(t)*. This step will be applied not only to *f(t)*, but also to the time processes obtained from *f(t)* by some sequential reductions of the spectrum of *f(t)*, by means of channel splitting. At the inputs of the new channels, the cutoff frequencies of these reductions will be defined on-line by parameters of the type $T_a$, found in the previous processing step. Since we use $P$ for finding $T_a$, at each application of the step we have the pairs $\{P, T_a\}$ that are just the points on the shape of the power spectrum. See [18] for a schematic of such sequential estimation of the compact spectrum, and see also [19] where a wide class of the some stochastic functions is shown to be relevant to the method.

We rejected the possibility of numerically estimating $E(t)$ because of a *clock* needed for sampling $f^2(t)$ in Riemann's construction of the integral. As was noted in Section 6.3, if the integration of $f^2(t)$ could be done according to Lebesgue's scheme, then, in principle, we could use $E(t)$, not introducing $\psi(t)$, for the determination of $T_a$. It would be interesting to consider technical realizations and the use of Lebesgue's scheme.

Regarding possible applications, one notes that since no clock is used, it can be assumed that not a lot of numerical calculations should be involved in the method, and thus the computation stage will be quickly done. The intention is to use as far as possible, analog devices. Such a method cannot be precise, but suitable for a quick preliminary estimation of spectrum, when the frequency range of the signal is not well known, which can mean that the signal appears unexpectedly. The associated point is that it may be not known whether the information contained in the signal is at all of interest for the receiver, and thus a very rough understanding the signal can be sufficient for deciding whether or not to analyze the signal using more precise and costly methods. These items can make the method relevant for communication between artificial intelligence systems; however permitting the side that initiates a contact by sending the signal to define the rate of the processing, or the basic time unit, is a gentlemanly behavior for any receiver.

Consider that the very concept of a function is introduced at an early stage of our mathematical education (in the secondary school) in a discrete manner, namely using a *table of values* representing the function, i.e. by means of a set $\{f(t_k)\}$, and this is always done with equidistant points. Sometimes this relates to a known function, e.g., $y = x^2$, that has to be drawn, and some points for the graph are needed, and sometimes this relates to a measurement of an initially unknown function like, say, temperature measured in shadow in a town, every three hours. It is important that even in the latter case one *knows* that for the actual variation-rate of the parameter being measured, the frequency of measurement/sampling is sufficient, i.e. nothing essential can be missed. This is also the situation of communication established between systems *which know each other*, and thus choose, for working with each other, the same frequency range.

However, if one does not sufficiently well know the range of the unknown input function/process, but applies the same method of the equidistant sampling, then one can easily have either an excessive sampling rate and too slow processing, or an insufficient sampling rate and some information missed. Thus, the use of equidistant sampling in the situation when the range of the signal is poorly known seems to be awkward, while relaying on the computer's power by always taking the maximal sampling rate (as permitted by the devices involved such as the sampler) seems to even be crude. "Sampling by definition", in the spirit of the Lebesgue's scheme, creates $f(t_k)$ for a more flexible set $\{t_k\}$ that adjusts itself to the rate of the changes of the input.

## 8. Conclusions and final remarks

Continuing the line of [3], the present work suggests that for some problems, presentation (1a) of a nonlinear system may be, at least heuristically, more useful than (1). This, of course, is not an objection to any known research directly based on (1), and for a problem where **F**(**0**,**u**,*t*) is nonzero, (1) can be preferable. However one sees that even when introduction of [A(**x**)] is analytically unnatural, the very argument regarding the nonlinearity of a system to be seen via (1a) remains in force, and, in general, the point of our outlook is first of all *logical*, and only then analytical. That is, the question is, first of all, whether or not the map **x** → **t*** *exists*, and only then what is it, while, in the present context, this "existence" is not just a formal mathematical problem, but something directly seen (defined) by the designer.

The similarity of the forms of (2) and (1a) leads to quite practical outlooks in the main sections and to some axiomatic points that are considered in detail in Appendix 2.

The overall list of the main points, including those in the Appendices is as follows.

1. The structural presentation of nonlinearity allows one to speak about a "nonlinear system" in terms of constructive definitions, associated with the state-equations. We thus speak about "*t*-systems", "**x**-systems", and "**u**-systems", of which the latter two relate to nonlinear systems. The (similar to the linear case) matrix-presentation of nonlinearity (1a) is strongly supported by the theory of switched systems which all are, first of all, some **t***(.)-systems. Form (1a) also seems to be more relevant than (1) as regards possible *generalizations* of nonlinear structures.

2. For the switched systems, **t***(.), i.e. either $f(t) \to$ **t***, or **x**$(t) \to$ **t***, is, generally, *a map* which is given in the operational terms, not as an analytical dependence or an explicit operator. In the case of **t***(**x**), some components of **x**$(t)$ are the inputs of the triggers, i.e. define the instants of singularity, as is illustrated by Fig. 1 (let there $f_1 \in$ **x**). This outlook should be clear for designer. However, e.g. [8], gives an example of another option to create **t***(**x**). In either case, **t*** are some numbers, i.e. **t***(**x**) is a functional. The present method of using level-crossings (i.e. the inequalities, '<', '>' between the levels around a $t_k$) better stresses that **t*** has to be a real valued vector; see [5] for a special discussion of this point.

3. Many switched systems can *not* be described *at the instants of switching* (see Appendix 1) as lumped systems, and the radiation at these instances has to be considered. The impossibility of including the radiation problem into the general theory of switched circuits is a serious lack of this theory. In the sense of positivistic philosophy (Appendix 2), these circuits are not sufficiently well theoretically described.

4. The numerous possibilities to introduce connections and, in particular, inputs, appearing in the field of modern electronic systems, suggest that we reconsider our general understanding of the differential equations. Namely, it is argued that "*a given input-function*" is not at all the same as "*a given function*" appearing in a description of the internal system operation. The input-function must be seen as belonging to a *set* of functions, and it should be *changed* in order to make clear the input-output map of the system. For instance, as the test of linearity shows, if $a(t)$ is the input of the associated system, then

$$\frac{dx}{dt} + a(t)x(t) = 0$$

is a nonlinear equation. An operator related to system description can be linear only if the functions included in it are properly fixed.

A simplified formulation, relevant to both **u**- and **u**-systems, might be here as follows.

*Checking the linearity of a system should not change this system; if the structure is changed, the system is nonlinear.*

5. Hydrodynamics gives a remarkable example of nonlinearity *in the sense of (1a)*, i.e. [A(**x**)]. The velocity field is both the unknown vector field to be found and the structure of the liquid system; thus, the [A(**x**)]-nonlinearity is inherent here. The nonlinear term in the Navier-Stokes equation,

$$(\vec{v}\nabla)\vec{v} ,$$

which is analogous to [A(**x**)]**x** in (1a), has the very basic origination from the "substation derivative"

$$\frac{d}{dt} = \frac{\partial}{\partial t} + (\vec{v}\nabla) ,$$

in which the nonlinear part cannot be eliminated by movement of the observer with the flow because $\vec{v}$ is not the same everywhere. Since thus, the stronger the vorticity, the stronger the nonlinearity is expressed, one sees that turbulence is caused by the nonlinear term, and that the turbulence and the degree of the nonlinearity enhance each other. Of course, linear terms of the equation should not cause turbulence, but we seek *direct arguments* for nonlinear effects.

One of the interesting questions here is whether or not the visible properties of a $(\vec{v}\nabla)$-system can be helpful in understanding an [A(**x**)]-system.

6. Regarding the linearization of the Navier-Stokes equation which was used in Section 5.1, and the more general interpretation of (2) as a linearized (1a), it needs to be noted that in technical literature *different ways* of linearization ("in average", in the sense of minimal r.m.s. error, etc.) of nonlinear characteristics and problems are found. Though after Lyapunov and Poincare, "linearization at a point" associated with stability theory (and also closest to elementary calculus concepts), is perhaps most well known, no general definition of linearization, *as a general method of coming from a nonlinear to a linear problem* exists.

7. If it can be legitimate to see an ensemble of colliding particles as a system in the circuit-theory sense, then, as we argue in Section 5.2, the chaos of the movement of the particles results from the system's nonlinearity which is of the **t\*(x)**-type. This observation touches different natural phenomena, and one can find additional examples of such a role of the **t\*(x)**-nonlinearity.

8. "Sampling by definition" is the simplest example of nonlinear sampling, and the nonlinear sampling is further considered using $\psi$-transform. Namely, we sample the nonlinear transform

$$\psi(t) = \int_0^t f^2(\lambda) sign[f(\lambda)] d\lambda$$

at the zero-crossings of the input process $f(t)$ allows one to develop a procedure for analyzing the spectrum of $f(t)$ without using any clock or time or frequency unit of measurement. The nonlinearity by $f$ of the set $\{\psi(t_k)\}$ used in Section 7 is a *double* one, -- that of the nonlinearity of $\psi(t)$ and that of the sampling of $\psi(t)$ at the zerocrossings of $f(t)$. Since, as one notes, taking higher powers in the integrands of both $P$ and $\psi(t)$ (or the associated "$E$") also leads to determination of some average period, the analytical component of the nonlinearity is not unique for the method, but the sampling at $\{t_k\}$ is such.

We find it relevant here to consider the nonlinearity of the approximating sums used in Lebesgues' scheme of construction of the integral, and also connecting this point with the elimination of a prescribed time-unit. Can one use these sums to obtain some independent nonlinear effects? Among the physical problems that might, perhaps, be relevant, the nonlinearity of the time-space metric in large-scale gravitational problems where the time axis/scale is connected with the spatial axis/scales and integrals can appear in calculations of total masses, can be noted.

9. From the position of engineering (i.e. reliable in applications) system theory, "nonlinear system" should be defined (Section A2.2) through structurally stable features of a system, which can be clearly observed. This rejects the definition of "nonlinear" as "not a linear one", because not passing a test of linearity is not yet evidence of clear and reliable nonlinear features, and simply because in order to become a really useful instrument, each concept has to be independently defined in physics terms. For instance, the two states of a trigger, which denote '1' and/or '0' have to be independently physically definable and reliably realizable. No engineer would accept, -- without independently checking *both* these modes, -- that 'not 1' is '0', and in a more general outlook, no reason exists for the particular case of binary logic with its however grammatically attractive use of the prefix "non" to be exception in the common rule of using in the tables of logical connections only signs/notations of some clearly defined objects or processes.

The requirement of structural stability is seen to be axiomatically important and, in this spirit, the effect chosen in [4] for observation of nonlinearity (the "Small Theorem") is a surely detectable effect.

10. In close connection with item 9, our position is that

$$\{t\text{-systems}\} \cap \{\mathbf{x}\text{-systems}\} = \varnothing,$$

(i.e. {LTV-systems}$\cap${NL systems} = $\varnothing$) and, in particular, for the respective subsets,

$$\{\mathbf{t^*}(t)\text{-systems}\} \cap \{\mathbf{t^*}(\mathbf{x})\text{-systems}\} = \varnothing,$$

which is because of the requirement of the structural stability of the nonlinear effects. Figuratively speaking, the areas of linear and nonlinear systems do not touch each other.

It follows, in particular, that LTI systems should not be interpreted as *both* a limiting case of {*t*-systems} *and* a limiting case of {**x**-systems}, when one might write

$$\{t\text{-systems}\} \cap \{\mathbf{x}\text{-systems}\} = \{\text{LTI -systems}\}.$$

11. The principle of constructiveness agrees with the positions of positivistic philosophy and intuinitionistic logic, and with practical and pedagogical reasoning. System theory is not just a branch of applied mathematics. However even from the mathematical point of view it has the right to its own axiomatization, as the consideration of **u**-systems as nonlinear systems, the argument of structural stability and some other arguments show.

## **Appendix 1: Some basic physical aspects of switched systems**

The present appendix discusses some *physical* specificity of switched circuits, which is important for understanding these circuits and their applications. We point out one advantage and one disadvantage.

### *A1.1. The advantage of good reliability*

An important advantage of electronic switched systems is the possibility to realize a strong nonlinearity while having a highly reliable circuit. We switch LTI elements, and the operational range of these elements, i.e. the range where they are very reliable, can be much wider than the range in which these elements are used for realizing the strongly nonlinear switching units.

This is distinct from the use of saturated inductors, or ferroelectric capacitors. Though use of analytical-nonlinearity will remain in basic micro and nano technology, such lumped elements hardly have a good feature in the industrial scheme-technique, and switched units are preferable there.

One has to see that the reliability problem regarding the "analytical nonlinearity" can prevent one from obtaining an acceptable "nonlinear circuit", because the official industrial acceptance of the "reliability" of an element can be a problem. The following example that the author met in his practice can be formulated as an "industrial theorem".

*It is impossible to produce, -- in the sense of a mass production that has to be done in a factory (not just in a laboratory), -- any circuit in which nonlinearity of ferroelectric capacitors is used.* Indeed, though as a rule, the given (specified) voltage range of ferroelectric ceramic capacitors ([25] and references there) is significantly lower than their breakdown voltage, the specification range is defined, by the producers, by the requirement of a fairly good linearity of the capacitors. The latter is natural since the main consumers are the electronics factories wishing to buy capacitors of a *certain capacitance*, i.e. linear capacitors.

Since the producers of any element always give data about the element only in the specified range, but electronics firm producing devices use elements only in the range for which several big producers of the elements give a specification with an associated guarantee of reliability, one cannot strongly exceed the linearity range in order to use the nonlinearity of these capacitors in a device intended for mass production.

One sees that the fact that a physical system does not pass a test of linearity in laboratory does not mean that we have a nonlinear "system" in the sense of *engineering, applicative* system theory.  The industrial reality influences the theoretical concepts and definitions, and students too have to be informed about that.

### *A1.2.  The radiation problem*

Any switching instant is a point of singularity, at which some of the components of the vector $d\mathbf{x}/dt$ do not exist.  However, the solutions of differential equations are defined on open sets where all components of $d\mathbf{x}/dt$ should exist.  Thus, at the points of switching we have to "sew" the local solutions of the equations, which are obtained in the intervals of analyticity, using some physical conditions.  The mathematical fact that we cannot speak about any differential equations at the instants of switching has an interesting physical meaning; we have *not*, generally, any *lumped system* described by Kirchhoff's equations, at these instants.

Assume first that we wish to increase a capacitance by connecting in parallel to a charged capacitor a capacitor having a different voltage, then the condition of conservation of electrical charge requires part of the electrostatic energy to be "lost" in the transient, which in the case of the really quick process *means radiation out of the circuit*.  Though lumped resistors can absorb some energy, at least part of the losses will be immediately radiated.  This situation contradicts the known condition [15] of sufficiently large wavelength needed for the lumped circuit description using Kirchhoff's equations, and Maxwell's field equations have to be used for the analyzing of the switching process.

The same situation is for switching of inductors when (e.g., because of the duality; $C \rightarrow L$ and $v \rightarrow i$) conservation of $q = Cv$ is replaced by conservation of $\psi = Li$, and just as $Cv^2/2 = q^2/(2C)$ is, generally, not conserved because $C$ is changed, also $Li^2/2 = \psi^2/(2L)$ is, generally, not conserved because $L$ is changed.

As a rule, the switching of *resistors* does not cause current pulses, but such a switching can make a current function discontinuous (e.g. when the resistor through which a capacitor is charged from a battery, is changed), i.e. the *derivative* of the current is pulsed.  This also is a singularity extending the frequency spectrum of the process, which can cause radiation, even if not so intensive as when the current by itself is pulsed.

It follows that switched circuits are "noisy" in the sense of the radiation and the resulted electromagnetic interference.  Though there are screening methods for reducing the electromagnetic interference, this is a disadvantage of these important circuits in the sense that *their theoretical description cannot be complete in principle*.

A similar remark relates to power-loss minimization problems for such a circuit. When trying to choose the optimal $v(q)$, or $i(\psi)$, or $v(i)$ characteristic of an element, which is intended to provide the circuit currents such that the power losses in the lumped resistors in the circuit would be reduced as much as possible, one may well increase the electromagnetic radiation by *thus-optimized* switched units.  Though the radiation is problematic and should be reduced, its power is usually much smaller than the power saved in the resistors because of the optimization.  Thus, when the electromagnetic interference is also the concern, there is no universal energetic criterion for circuit-optimization in switched circuits.

It is interesting to note, however, that the radiation can be a good indication of switching in a system (see [4]), which presents switching as a clearly registerable phenomenon. This is a side-effect, but a positive aspect of the radiation problem.

One can compare the switched circuit with a quantum system, in which the instants of emission or absorption of a photon separate between well-describable stationary states of the system. Another comparison can be with a mechanical system, as, e.g. that in Section 5.2.

## Appendix 2: On the basic concept of nonlinearity and the requirement of constructive definitions in system theory

The concept of linearity is defined purely mathematically, and even when not yet deepening, as below, on the very important aspect of *constructiveness*, but just requiring, for seek of justice, the concept of nonlinearity to also be independently mathematically defined, one already comes through (1a) to the notations "**x**-system" (or "**t\*(x)**-system") and "**u**-system". However these notations, characterizing different nonlinear systems can/should hardly replace the very concept of *nonlinearity*, which is a "collecting concept", *useful for our intuition*, and we only can stress here the importance of the constructive aspect, not directly presented in this concept. The stress on constructiveness is seen to be more necessary when one notes that system theory needs more constructive formalizations in some other fields as well, e.g. in the field of infinite systems.

### A2.1. Some comments on the definition "nonlinear is not a linear one"

The "grammatical" definition, which, in its simplest form, is

$$\text{"}nonlinear\ is\ not\ a\ linear\ one\text{"} \qquad (A1)$$

and in [26] "The circuit is called *nonlinear* iff it is *not linear*", is non-constructive, since an unknown is defined via something not known. In order to see what (A1) is about, one has to know that circuits/systems other than linear ones exist at all. For this, some constructive examples of nonlinear systems have to be known to one before (A1). In other words, for (A1) to be understood, one already has to know what "nonlinear system" is.

It is hardly desirable to connect the concepts of linear and nonlinear systems at all because of the very different features of these systems (one of the concepts does not help one, rather the opposite, to think in terms of the other, because of the very different features of the different systems), but if one wishes, one can use the statement:

$$\text{"}linearity\ is\ a\ particular\ case\ of\ nonlinearity\text{"}. \qquad (A2)$$

Indeed, one can consider (see Fig.6) in a plane, or 3D space, any curve, here representing "nonlinearity", and a straight line, here representing the linearity, asking the question of which is more correct, -- to say that the curve is "not a straight line", or that the straight line is a particular case of the curve?

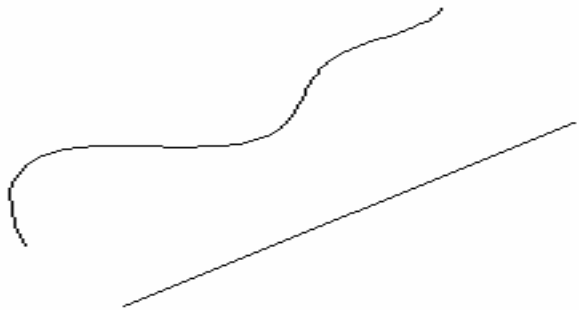

Fig.6: Which line is initial?

From the comparative geometrical point of view and also from the positions of the variational calculus, a straight line (linearity) is just the particular/special case of *the curves* (nonlinearity) -- that which minimizes the curve's length between two given points. It thus seems that the concept of the curve is more general and need not be defined using the concept of the straight line.

Quite similarly, it is obvious that (1a) is more general than (2), and when [A($t$,**x**)] and [B($t$,**x**)] in (1a) are independent of **x**, or become known/prescribed by substituting in them a certain known approximation/estimation of **x**, we obtain a linear system, as a special case of (1a). Certainly, one should not define (1a) as "not (2)".

Of course, the analogy between the lines in Fig. 6 and equations (1a) and (2) not only simplifies the situation. The definition of line in topology is problematic [26,27], and for one the analogy of Fig. 6 may even raise the question of whether or not we can write state-equations for a very complicated physical system. This comment, however, does not increase the importance of linear systems here, and is not practical regarding the electrical systems that can come into the consideration. It is clear that any, not too complicated, curve can be built (drawn) without any definitional treatment.

On the psychological regard, one notes, however, that if one has in sight many drawn linear pieces and only one drawn curve, *then* it is natural for one to call the curve "not straight line". The latter situation is the real one since one usually knows much more about linear systems, and the concept of linear system thus naturally becomes the "logical reference". Since we live in weak fields, apply weak forces, which all lead to linearizations, and also like sinusoids that are associated with simple rotational movement, it was natural and mathematically easier to start the sciences from linear systems. However, this situation is changing. Because of the wide use of electronic switched systems, nonlinear systems have become very common, perhaps more common than linear, and this should influence the interests and education of future engineers and circuit specialists. According to these arguments and the mentioned position of [1], nonlinear system-theory should be founded independently.

### A2.2. *The requirement of structural stability of the operation of a system.*

Since nonlinear science is partly empirical, even such a special case where a system fails a test of linearity *just because it is faulty*, or exploded, cannot be ignored in the criticism of (A1).

Not every *physical system/object* is a "system" in the sense of circuit theory, and the problem of having a nonlinear system may be, in a particular case, the problem of having a "system" at all. Thus, any definition of "nonlinear system" must be subjected to a general definition of "system" which would be accepted in engineering circuit theory.

Perhaps the most practical reason explaining why *if a system does not pass a test of linearity (i.e. is "not a linear one") this does not yet mean that it can be considered as "nonlinear" and operated as such*, is given by the basic requirement of engineering system theory to deal only with those systems whose features are surely (structurally stably) measured. This requirement relates, of course, to a system taken as a whole, i.e. to its overall application side. A chaos generator can unstably undergo numerous bifurcations of its internal dynamics which leads to the chaotic attractor, but this very attractor has to be stably observed for the system to be reliably used as the chaos generator, i.e. the *statistical properties* of the attractor (or chaos) should be stable. Recall also the argument of Section A1.1 associated with the problem of reliability.

Since not passing a test of linearity does not ensure structural stability of any nonlinear features, one can schematically imagine the fields of linear and nonlinear systems as some geometric areas not touching each other. Statement (A1) requires the systems in the "separating area" to be classified as "nonlinear systems", but the physical systems belonging to the separating area at all should not be considered as "systems" in the sense of the applicable system theory.

### *A2.3. The axiomatic side and a historical view*

The requirement of the constructiveness of scientific definitions is not new, and the "intuitionistic" logic/mathematics" by L. Brouwer [29,30] is even centered on such a requirement. Namely, it is assumed that no "*A*" can be proved/obtained by rejection of "not *A*", i.e. the very common *tertium non datur* ("third not given") principle saying that the disjunction $A \cup \bar{A}$ is *complete*, is not accepted in the intuitionistic logic.

However, the basic point here is *not* that one thinks that there can be something *third*, not *A*, and not $\bar{A}$, but the opinion/belief that *rejection of one thing cannot prove the **existence** of another thing, -- a direct **intuitively convincing** example of **construction** of the latter thing is needed*. In terms of probabilities, the commonly used equality

$$P\{A \cup \bar{A}\} = 1, \qquad (A3)$$

*taken by itself*, is rejected if at least one of the symbols, *A*, or $\bar{A}$, is not clearly constructively defined. In the context of the definition of nonlinearity, this means that it is insufficient to know that the test of linearity is not passed in order to have a *construction* that might be seen as a nonlinear *system*.

Since the development of computers through the performance of the arithmetical operations did not require that intuitive arguments be involved, the intuitionistic logic, developed at the beginning of the past century, "lost" the "competition" against the usual logic that operates, in particular, Boolean algebra. *However, from the empirical point of view, intuitionism agrees, in its basic point with the usual logic*. Indeed, the success of the usual logic in the field of computers is based on the *reliable technical realization* of 1 and 0 ("truth", "false", *both* constructively well-defined). If non-realization of a certain state of a flip-flop did not technically mean certain realization of another state (if not '1', then surely '0', and opposite), -- then engineers too would be against the use of the *tertium non datur* principle and Boolean algebra. The order of the things is such that one first fills a table with the notations (as '1' and '0') of some *existing physical states that can be reliably used*, and only then, if there are only two such states, can make such conclusions as 'not 1' is '0', etc.. However successful in the applications (and simply grammatically formulated using the prefix "non"), the binary case is just a particular one here, and nothing can be changed in this basic order/scheme.

Thus, in its rejection of non-constructive objects or procedures, intuitionism tried to extend *the same principle* of the good "technical realizability" that was adopted by conventional logic, to other branches of mathematics. However, being occupied with mathematical axioms, Brouwer and, e.g., the authors of [30] could not express themselves in such convincing physical terms.

The role of switched systems in the main text is not occasional also regarding the intuitionistic logic, because switchings are easily registerable/measurable parameters (again, the Small Theorem of [4] is an example.) It is interesting to observe that in the history of science, switching systems already played important role when Shennon used, after almost 100 years since the Boolean algebra appeared, this algebra for creating his theory of switched circuits that lead to the development of computers.

More about intuitionistic logic can be found in [30], and more about Brouwer's overall contribution to mathematics (including the proof of the fixed-point theorem) and personality in [31,32].

Regarding the historical background, it should be also noted that the requirement of constructivism itself also comes from the much earlier, and more widely known than the intuitivism, "positivistic philosophy" [29] supported, in particular, by Kirchhoff, which must have influenced Brouwer. According to positivist philosophy, one always has first to learn how to *well* (*completely and as simply as possible*) *describe* things and only then how to *explain* them, so that the primary goal of science is *description* of natural phenomena, which obviously requires constructivism.

Of course, constructivism is relevant to the purely theoretical treatments too; Kirchhoff would not be satisfied by the existing theory of switched circuits because of the lack of description of the radiation at the switching instants, as is considered in Section A1.2.

The positivistic philosophy had been proved important each time when new sciences started and when a *difficult field* (e.g., thermodynamics, or quantum mechanics) had been developed. The field of nonlinear systems seems to be sufficiently difficult for the positivistic approach to be acceptable for it, and even a theoretician can accept that the *physical existence of things is more important than their theoretical explanations*. Statement (A1) is very far from any such constructive thesis.